\documentclass[twocolumn,showpacs,amsmath,amssymb,floatfix,prb,preprintnumbers,footinbib]{revtex4}

\usepackage[dvips]{graphicx}
\usepackage{dcolumn}% Align table columns on decimal point
\usepackage{bm}% bold math
\usepackage{amsfonts}

\begin{document}

\title{Signatures of spin-related phases in transport through regular polygons}  

\author{Dario Bercioux}\affiliation{Institut f{\"u}r Theoretische
Physik, Universit{\"a}t Regensburg, D-93040, Germany \\
Coherentia-INFM \& Dipartimento di Scienze Fisiche,
Universit\`a degli Studi ``Federico II'', I-80126 Napoli, Italy}

\author{Diego Frustaglia}
\author{Michele Governale}  
\affiliation{NEST-INFM \& Scuola Normale Superiore, I-56126 Pisa, Italy}

\date{\today}

\begin{abstract}
We address the subject of transport in one-dimensional ballistic polygon 
loops subject to Rashba spin-orbit coupling. We identify the role played by the polygon 
vertices in the accumulation of spin-related phases by studying interference effects
as a function of the spin-orbit coupling strength. We find that the vertices act as strong 
spin-scattering centers that hinder the developing of 
Aharovov-Casher and Berry phases.
In particular, we show that the oscillation frequency of interference pattern can be 
doubled by modifying the shape of the loop from a square to a circle.
\end{abstract} 


% 03.65.Vf Phases: geometric; dynamic or topological
% 71.70.Ej Spin-orbit coupling in condensed matter
% 73.23.-b Electronic transport in mesoscopic systems
\maketitle

Many efforts have been done towards the study of spin effects at 
the mesoscopic scale since the original Datta-Das proposal \cite{DD90} 
for a spin-field effect transistor. 
This is based on the control of the Rashba spin-orbit (SO) coupling~\cite{rashba-1960} 
in low-dimensional electron gas subject to asymmetric quantum confinement.
As a consequence, a large variety of alternative setups relying on similar 
principles has been presented.  
Systems of particular interest are
the spin interferometers~\cite{ACrings,frustaglia-2004} and their
extension to quantum networks~\cite{bercioux-2004} which do not
require the injection of spin-polarized carriers as a working
principle. 
Instead, they work by only tuning the Aharonov-Casher (AC) phase\cite{AC84} 
acquired by spin carriers in the presence of SO coupling.
Spin interferometers are interesting not only in view of possible
spintronics~\cite{ZFDS04} applications but also from a fundamental
perspective regarding the study of spin dynamics and related quantum
phases.
For instance, conducting rings have been 
proposed~\cite{theoBP} as
paradigmatic systems for the identification of geometrical or Berry
phases~\cite{B84} with relative experimental
success~\cite{expBP,YYLG04}. Berry phases arise when the spins suffer
an adiabatic evolution during transport, \emph{i.e.}, when they adiabatically
follow the local direction of the effective magnetic field during
transport (see e.g. Ref.~\onlinecite{frustaglia-2004}). Among the
different factors that can affect the spin evolution (and consequently
the eventual presence of Berry phases) it stands out disorder, a
subject extensively discussed in the literature~\cite{PFR03}.
Surprisingly, less interest has been put 
on other relevant geometrical aspects.  
Only recently Yang \emph{et al.}~\cite{YYLG04} proposed
alternative setups optimizing the contacts to the leads in order to
avoid eventual non-adiabatic spin flipping, and van Veenhuizen
\emph{et al.}~\cite{VKN04} discussed single-probe spin-interference
features in closed polygon-structures.  

In this Brief Report we discuss the two-contact transport properties
of regular polygons subject to Rashba SO coupling. 
We approach the
subject 
by means of the Landauer-B\"uttiker\cite{landa-butt} formulation, 
identifying the linear conductance with
quantum transmission. 
We calculate the linear conductance of
several polygons made of one-dimensional (1D) ballistic (disorder
free) wires as a function of the SO strength.  As a result we obtain a
series of interference patterns reproducing the characteristics of the
spin-related phase accumulation. This permit us to identify the role
played by non-adiabatic spin-scattering taking place at the vertices
of the polygons.
%
%
%%%%%%%%%%%%%%%%%%%%%%%%%%%%%%%%%%%%%%%%%%%%%%%%%%%%%%%%%%%%%%%%%%%%%%%%%%%%%%%%%%%%
%					FIGURE
%%%%%%%%%%%%%%%%%%%%%%%%%%%%%%%%%%%%%%%%%%%%%%%%%%%%%%%%%%%%%%%%%%%%%%%%%%%%%%%%%%%%
\begin{figure}
	\centering \includegraphics[width=3.0in]{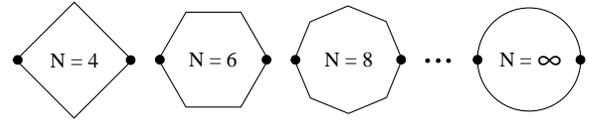}
	\caption{Series of regular-polygon conductors of constant perimeter. 
	Vertices are connected by single-channel ballistic quantum wires with SO 
        coupling. In the limit of infinite number of vertices the series converges to 
	a single-channel circular conductor. The full dots represent the point where input 
	and output leads are attached. \label{polygones}}
\end{figure}
%%%%%%%%%%%%%%%%%%%%%%%%%%%%%%%%%%%%%%%%%%%%%%%%%%%%%%%%%%%%%%%%%%%%%%%%%%%%%%%%%%%%
%

We consider electron transport through regular polygons with an even
number of vertices $N$ as those shown in Fig.~\ref{polygones}. The
vertices are connected by single-channel ballistic quantum wires subject
Rashba SO coupling.  The polygons are symmetrically coupled to two 1D
leads free of SO coupling at opposite vertices.  This model neglects
the subband hybridization due to the Rashba
effect~\cite{mir-gov}. (See
Ref.~\onlinecite{bercioux-2004} for a related model). The Hamiltonian 
for a single-channel wire along a generic direction $\hat{\gamma}$ in
the $x$-$y$ plane reads
%
%%%%%%%%%%%%%%%%%%%%%%%%%%%%
\begin{equation}
\label{hamiltonian}
	\mathcal{H} = \frac{p_{\gamma}^2}{2\,m} -
	\frac{\hbar k_{\text{SO}}}{m}p_\gamma\left(
\hat{z} \times \hat{\gamma} \right)\cdot\vec{\sigma},
\end{equation}
%%%%%%%%%%%%%%%%%%%%%%%%%%%%
%
where $k_{\text{SO}}$ is the SO coupling strength, and $\vec{\sigma}$
is the vector of the Pauli matrices. 
The second term in Eq.~(\ref{hamiltonian}) can be viewed as a Zeeman splitting 
due to a momentum-dependent, in-plane, effective magnetic field. 
 The SO coupling strength
$k_{\text{SO}}$ is related to the spin precession length
$L_{\text{SO}}$ by $L_{\text{SO}}=\pi/k_{\text{SO}}$. For InAs
quantum wells the spin-precession length ranges from $0.2$ to $1$
$\mu$m~\cite{NSG}. 

The spin dynamics along one side of a polygon can be described by a 
spin-rotation operator (SRO)
that accounts for spin precession around the effective in-plane magnetic
field due to the SO coupling \cite{bercioux-2004}:
%
%
%%%%%%%%%%%%%%%%%%%%%%%%%%%%
\begin{equation}
\label{rotation}
 	\mathcal{R}_{q,p}(k_{\text{SO}}) = \exp\left\{-i
 	\vec{\sigma}\cdot \left( \hat{z} \times \hat{\gamma}_{q,p}
 	\right) k_{\text{SO}} l_{q,p} \right\},
\end{equation}
%%%%%%%%%%%%%%%%%%%%%%%%%%%%
%
%
where $\hat{\gamma}_{q,p}$ and $l_{q,p}$ are the orientation and
length of the bond connecting the vertices $p$ and $q$,
respectively. The vertices are numbered clockwise from 1 to $N$ ($N$
even). The incoming (outgoing) lead is coupled to vertex 1
($N/2+1$). For regular polygons it is $l_{q,p}=P/N \equiv l_N$
independently of the particular vertices involved, where $P$ is the
perimeter of the polygon.
%
%%%%%%%%%%%%%%%%%%%%%%%%%%%%%%%%%%%%%%%%%%%%%%%%%%%%%%%%%%%%%%%%%%%%%%%%%%%%%%%%%%%%
%					FIGURE
%%%%%%%%%%%%%%%%%%%%%%%%%%%%%%%%%%%%%%%%%%%%%%%%%%%%%%%%%%%%%%%%%%%%%%%%%%%%%%%%%%%%
\begin{figure}
	\centering \includegraphics[width=2.7in]{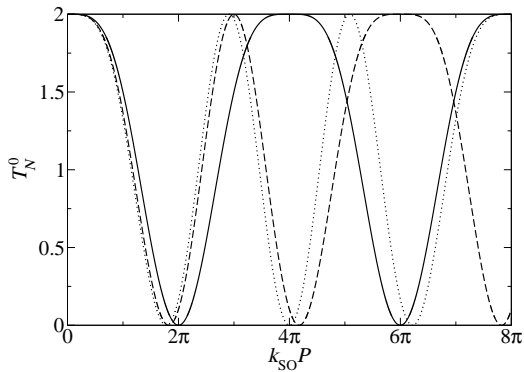}
	\caption{\label{fig2} Lowest order transmission as function of
	the dimensionless SO coupling strength for polygons with $N=4$ (solid line), 
	$N=6$ (dashed line) and $N=8$ (dotted line).}
\end{figure}
%%%%%%%%%%%%%%%%%%%%%%%%%%%%%%%%%%%%%%%%%%%%%%%%%%%%%%%%%%%%%%%%%%%%%%%%%%%%%%%%%%%%

We study the spin-dependent conductance of polygons
by employing two different methods: A full-quantum-mechanical (FQM)
approach, 
based on quantum graph theory~\cite{bercioux-2004}, and
a much simpler description  
that accounts only for the spin
dynamics at the lowest order in orbital winding by using the SRO of
Eq.~(\ref{rotation}).  The latter approach provides the spin-related
AC phases accumulated by the carriers between input and output leads by
following the two shortest possible paths (namely, the direct
clockwise and counterclockwise paths neglecting any further winding and
backscattering contributions). The corresponding quantum amplitude for
transmission is calculated as the sum of successive application of the
operator (\ref{rotation}) along both paths, each of them composed by
$N/2$ 1D conducting bonds.  This reads
%
%
%%%%%%%%%%%%%%%%%%%%%%%%%%%%
\begin{eqnarray}\label{amplitude}
	\Gamma_N^0 &=& \mathcal{R}_{N/2+1,N/2} \ldots
	\mathcal{R}_{3,2} \cdot \mathcal{R}_{2,1}+ \\  
	&~& \mathcal{R}_{N/2+1,N/2+2} \ldots \mathcal{R}_{N-1,N} \cdot
	\mathcal{R}_{N,1}, \nonumber 
\end{eqnarray}
%%%%%%%%%%%%%%%%%%%%%%%%%%%%
%
where $\Gamma_N^0$ is a $2 \times 2$ matrix containing the
spin-resolved amplitudes and the index $0$ stands for
lowest-order-contribution. The first and second terms in
Eq.~(\ref{amplitude}) concern the clockwise and counterclockwise
paths, respectively.
The associated transmission coefficient (proportional to the
linear conductance) 
is given by %
%
%%%%%%%%%%%%%%%%%%%%%%%%%%%%
\begin{equation}
\label{transmission}
	T_N^0 = \text{Tr}\left[\Gamma_N^0 \Gamma_N^{0 \dag} \right],
\end{equation}
%%%%%%%%%%%%%%%%%%%%%%%%%%%%
%
%
where the trace runs over the spin degree of freedom.
In Fig.~\ref{fig2} we plot $T_N^0$ as a function of the 
dimensionless SO strength
$k_{\rm SO} P=\pi P/L_{\rm SO}$ for $N=4,~6$ and $8$. We find a
series of zeros showing up due to destructive quantum
interference. This stems from the fact that spins following different
paths acquire different 
AC phases according to the traveling
direction~\cite{ACrings,frustaglia-2004,bercioux-2004}.  The curves
present a periodicity 
equal to $N \pi$, proportional to the number of 
vertices. The  
number of zeros within a period equals $N/2-1$. For $N >
4$ it becomes evident the presence of two very different frequencies
participating in the oscillatory pattern.  
The first one, associated
to the shorter bond-length-scale $l_N$, determines the absolute period
of the curves increasing linearly with $N$ as pointed out above.  The
second frequency is much higher and weakly dependent on $N$. This is
related to the longer perimeter-length-scale $P$, giving rise to
oscillations of period between $2 \pi$ and $3 \pi$. 
As a function of $N$, it approaches $2 \pi$ as $N$
increases.
%
%%%%%%%%%%%%%%%%%%%%%%%%%%%%%%%%%%%%%%%%%%%%%%%%%%%%%%%%%%%%%%%%%%%%%%%%%%%%%%%%%%%%
%					FIGURE
%%%%%%%%%%%%%%%%%%%%%%%%%%%%%%%%%%%%%%%%%%%%%%%%%%%%%%%%%%%%%%%%%%%%%%%%%%%%%%%%%%%%
\begin{figure}
	\centering 
	\includegraphics[width=2.7in]{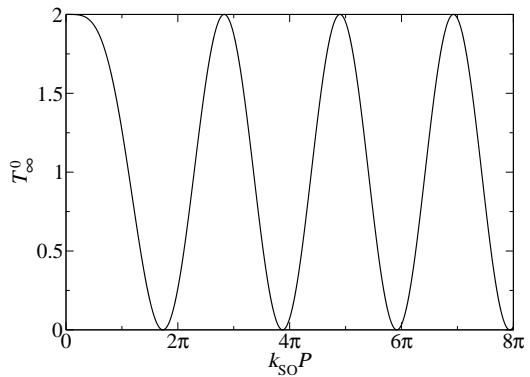}
	\caption{\label{fig3} Lowest order transmission as function of
	the dimensionless SO coupling strength for a polygon in the limit $N \rightarrow \infty$,
	equivalent to a circular ring.}
\end{figure}
%%%%%%%%%%%%%%%%%%%%%%%%%%%%%%%%%%%%%%%%%%%%%%%%%%%%%%%%%%%%%%%%%%%%%%%%%%%%%%%%%%%%
%

These features are better understood by taking the limit $N
\rightarrow \infty$, where the series of regular polygons converges to
a circle (Fig.~\ref{polygones}). Results for $T_N^0$ in the limit of
large $N$ ($T_\infty^0$) are shown in Fig.~\ref{fig3}, where we recover the results
for a ring recently reported by Frustaglia and Richter in 
Ref.~\onlinecite{frustaglia-2004}. 
There, we find that the series of
zeros in Fig.~\ref{fig3} are placed at\cite{note1}
%
%
%%%%%%%%%%%%%%%%%%%%%%%%%%%%
\begin{equation}\label{diegoszero}
	k_{\text{SO}} P = \pi~\sqrt{(2n)^2-1}, 
\end{equation}
%%%%%%%%%%%%%%%%%%%%%%%%%%%%
%
%
with $n$ integer. In contrast to the case of polygons, only higher
frequency oscillations associated to the perimeter length scale show
up. These appear to be quasi-periodic, with a period approaching $2 \pi$ 
as $k_{\rm SO} P \rightarrow \infty$. The latter
corresponds to the adiabatic limit, where spins follow the local
direction of the effective in-plane magnetic field during transport, and Berry
phases arise~\cite{frustaglia-2004}. Oscillations of period $2 \pi$ can
consequently be identified with the adiabatic limit. From our
discussion regarding results for finite $N$ in Fig.~\ref{fig2} (see
above) we conclude that the limit of adiabatic spin transport is never
really achieved in polygons, and it may be only approached for large
$N$.  This is due to the strongly non-adiabatic scattering that the
spins suffer at the vertices
as a consequence of the abrupt change of direction 
of the effective magnetic field. 
That is particularly
relevant in square loops ($N=4$) where the period of the oscillations
doubles that for rings (compare Fig.~\ref{fig2} (solid line) with
Fig.~\ref{fig3}). This indicates that the rate of AC phase accumulation as a function 
of $k_{\rm SO} P$ is smaller for the square loop. 

%%%%%%%%%%%%%%%%%%%%%%%%%%%%%%%%%%%%%%%%%%%%%%%%%%%%%%%%%%%%%%%%%%%%%%%%%%%%%%%%%%%%
%					FIGURE
%%%%%%%%%%%%%%%%%%%%%%%%%%%%%%%%%%%%%%%%%%%%%%%%%%%%%%%%%%%%%%%%%%%%%%%%%%%%%%%%%%%%
\begin{figure}
	\centering \includegraphics[width=2.7in]{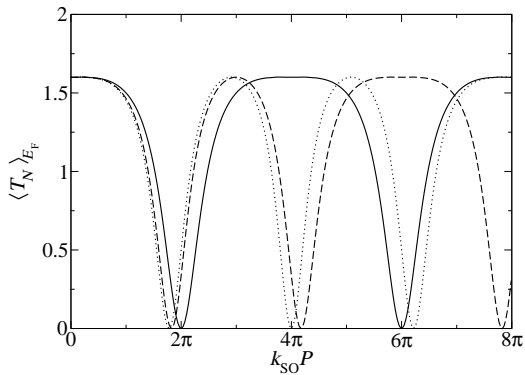}
	\caption{\label{fig4} Full average transmission as function of
	the dimensionless SO coupling strength for polygons with $N=4$ (solid line), 
	$N=6$ (dashed line) and $N=8$ (dotted line). To be compared
	with the lowest order results of Fig.~\ref{fig2}.}
\end{figure}
%%%%%%%%%%%%%%%%%%%%%%%%%%%%%%%%%%%%%%%%%%%%%%%%%%%%%%%%%%%%%%%%%%%%%%%%%%%%%%%%%%%%
%
We finally present results using a FQM approach which takes into
account higher order winding contributions and backscattering effects
due to finite coupling to the leads~\cite{bercioux-2004}. Results are
shown in Fig.~\ref{fig4} for polygons with $N=4,~6$ and $8$ and are to
be compared with those of Fig.~\ref{fig2} for the SRO approach.  There
we plot the corresponding average transmission $\langle T_N
\rangle_{E_{\rm F}}$.  
The average is performed on the Fermi energy
$E_{\rm F}$ for a energy window larger than the mean level spacing. 
This is done in order to avoid energy-dependent features related to the presence 
of quasibound states 
which are formed in the 
polygons when the coupling 
to the leads is finite (see Ref.~\onlinecite{frustaglia-2004} for a similar
procedure). By comparing Fig.~\ref{fig4} with Fig.~\ref{fig2} we note that both
approaches provide the same set of points of zero conductance. The
main differences are in the shape and amplitude of the curves. 
This is
due to the fact that in the simpler SRO approach we assumed an ideal
coupling to the leads, 
and we consider only the lowest order terms in orbital winding.  
Higher order contributions, due to paths that 
go several times around the polygons, modify slightly the shape of the curves.  
Regarding the amplitude
reduction, this is 
mainly due to backscattering at the incoming contact. 

In conclusion, we studied the spin-dependent transport properties of 1D polygons
subject to SO coupling. We showed that the polygons vertices act as scattering
centers for spin, leading to highly non-adiabatic spin 
evolution.
This hinders the spins 
to reach the limit of adiabatic spin transport where Berry phases manifest. For polygons 
with large number of vertices the adiabatic regime is restored for large coupling strength,
in agreement with previous results for circular rings. It is interesting to note
that by simply changing the shape of the loop from a square to a circle one can double 
the frequency of the oscillation pattern as a function of the 
dimensionless SO coupling strength.  
Moreover, we remark that a simple approach based on SRO captures the relevant 
spin dynamics, giving the exact positions of the conductance zeros. 
We finally note that our results are not expected to suffer any significative change in 
the presence of residual disorder, as far as the elastic mean free path stays of the 
order of the semi-perimeter $P/2$ (i.e., no new length scales are introduced). 
Some effects would evetually manifest on the amplitude of the oscillations, where the 
minima would not reach zero due to the symmetry breaking introduced by disorder. 

{\it Acknowledgments:} This work was partly supported by the EU Spintronics Research 
Training Network (DF).

%%%%%%%%%%%%%%%%%%%%%%%%%%%%%%%%%%%%%%%%%%%%%%%%%%%%%%%%%%%%%%%%%%%%%%%%%%%%%%%%%%%%
%				   BIBLIOGRAPHY
%%%%%%%%%%%%%%%%%%%%%%%%%%%%%%%%%%%%%%%%%%%%%%%%%%%%%%%%%%%%%%%%%%%%%%%%%%%%%%%%%%%%

%%%%%%%%%%%%%%%%%%%%%%%%%%%%%%%%%%%%%%%%%%%%%%%%%%%%%%%%%%%%%%%%%%%%%%%%%%%%%%%%%%%%

\end{document}